\documentclass[reprint, amsmath, amssymb, aps, superscriptaddress]{revtex4-1}
\usepackage{format}
\allowdisplaybreaks

\begin{document}

\preprint{APS/123-QED}

\title{ Viscous damping in weltering motion of trapped hydrodynamic dipolar Fermi gases }

\author{Reuben R. W. Wang}
\author{John L. Bohn}
\affiliation{ JILA, NIST, and Department of Physics, University of Colorado, Boulder, Colorado 80309, USA }
\date{\today} 

\begin{abstract}

We consider collective motion and damping of dipolar Fermi gases in the hydrodynamic regime.
We investigate the trajectories of collective oscillations -- here dubbed ``weltering'' motions -- in cross-dimensional rethermalization experiments via Monte Carlo simulations, where we find stark differences from the dilute regime. These observations are interpreted within a semi-empirical theory of viscous hydrodynamics for gases confined to anisotropic harmonic potentials. 
The derived equations of motion provide a simple effective theory that show favorable agreement with full numerical solutions. To do so, the theory must carefully account for the size and shape of the effective volume within which the gas' behavior is hydrodynamic.
Although formulated for dipolar molecules, our theoretical framework retains a flexibility to accommodate arbitrary elastic cross sections.

\end{abstract}

\maketitle

\section{ \label{sec:introduction} Introduction }

Suppression of two-body collisional losses has been crucial for achieving stable samples of molecular quantum gases.  
Within the last decade, theoretical and experimental advances have brought to fruition the electric field shielding of polar molecules against chemical reaction and complex formation \cite{Karman18_PRL, Karman20_PRA, Anderegg21_Sci, Bigagli23_arxiv, Quemener16_PRA, Gonzalez17_PRA, Matsuda20_Sci, Li21_Nat, Lassabliere22_PRA, Bigagli23_arxiv, Lin23_arxiv}, permitting the production of degenerate bulk molecular samples \cite{Valtolina20_Nat, Schindewolf22_Nat}. 
But even before the onset of quantum degeneracy, these shielded molecules present a long-lived versatile platform for exploring dipolar physics \cite{Baranov08_PR, Lahaye09_IOP, Chomaz22_IOP}. For instance, dipole-dipole interactions lead to highly anisotropic two-body collision cross sections \cite{Bohn14_PRA} and observable anisotropy in the collective dynamics of thermal gases \cite{Aikawa14_PRL, Tang16_PRL, Sykes15_PRA, Wang20_PRA, Wang21_PRA}. 
For these nondegenerate bulk gases, thermalization is an essential mechanism with great utility in applications such as evaporative cooling \cite{Anderson95_Sci, Davis95_PRL, DeMarco99_Sci, Thomas03_JOB, Griesmaier05_PRL, Lu11_PRL, Aikawa12_PRL, Marco19_Sci} and scattering length measurements \cite{Monroe93_PRL, Newbury95_PRA, Schmidt03_PRL, Tang15_PRA, Patscheider21_PRA}. 
The accuracy and efficacy of both these applications, in turn, rely on a deep understanding of thermalization in such systems.

\begin{figure}[ht]
    \centering
    \includegraphics[width=0.95\columnwidth]{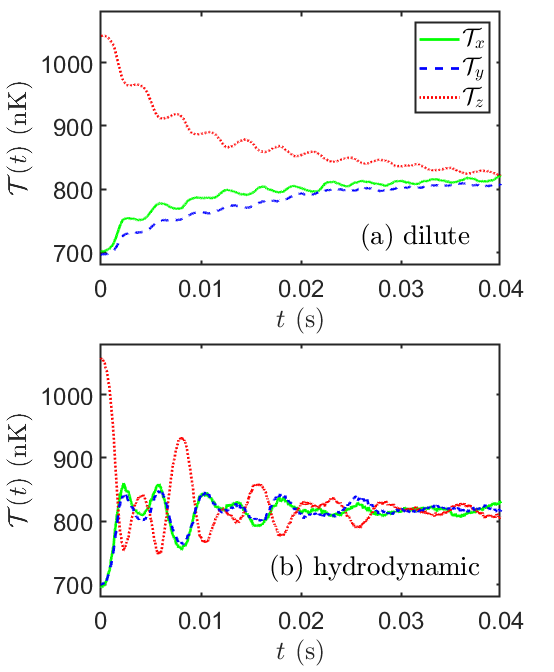}
    \caption{ Pseudotemperatures (\ref{eq:pseudotemperature}) obtained from Monte Carlo simulations in the dilute (upper panel, a) and hydrodynamic (lower panel, b) regimes. 
    The gas consists of microwave shielded $^{23}$Na$^{40}$K molecules with dipole moment $d = 0.75$ D, oriented along $\hat{\boldsymbol{x}}$, at temperature $T = 700$ nK. 
    The gas is initially excited along $z$ by an instantaneous trap frequency ramp to $\omega_z = 2 \pi \times 147$ Hz, while $\omega_x = \omega_y = 2 \pi \times 82.5$ Hz remain constant.
    The regimes are differentiated by the number of molecules $N$, which are $N = 10^4$ in panel (a), and $N = 2 \times 10^5$ in panel (b). }
    \label{fig:pseudotemps_DLvsHD_Theta90}
\end{figure}

The difference between dilute and hydrodynamic limits is revealed clearly in a gas' response to perturbation.  In particular, in a cross-dimensional rethermalization experiment, an initially equilibrated gas is preferentially heated along a particular axis, then allowed to rethermalize back to equilibrium \cite{Monroe93_PRL}. 
Thermalization in the dilute regime is closely related to the collision rate \cite{Monroe93_PRL, Guery99_PRA, Wang21_PRA}, while the hydrodynamic regime sees similarly extracted relaxation rates close to the trapping frequency instead \cite{Ma03_JPB, Thomas03_JOB, Schindewolf22_Nat}. 
The difference between the two regimes is illustrated in  Fig.~\ref{fig:pseudotemps_DLvsHD_Theta90}.  In both panels, a collection of $^{23}$Na$^{40}$K molecules is subjected to the same harmonic trapping potential
\begin{align} \label{eq:total_potential}
    V(\boldsymbol{r}) 
    &
    = \frac{1}{2} m
    \sum_i \omega_i^2 r_i^2.
\end{align}
and subsequently excited along the $z$ axis.  The only difference is the molecule number: for fewer molecules in the upper panel (a), the dynamics is dilute, while for a greater number of molecules in the lower panel (b), it is hydrodynamic.

In both cases, the behavior is tracked using time trace plots of the pseudotemperatures ${\cal T}_i(t)$, shown in Fig.~\ref{fig:pseudotemps_DLvsHD_Theta90}. A pseudotemperature is defined along axis $i$ as \cite{Sykes15_PRA}
\begin{align} \label{eq:pseudotemperature}
    k_B {\cal T}_i(t)
    \equiv 
    \frac{ 1 }{ 2 }
    m \omega_i^2 \{ r_i^2 \}(t)
    +
    \frac{ 1 }{ 2 }
    m \{ v_i^2 \}(t).
\end{align}
where $\{ \ldots \}(t)$ denotes the time varying ensemble average over molecular positions $\boldsymbol{r}$ and velocities $\boldsymbol{v}$, $m$ is the molecular mass, $k_B$ is Boltzmann's constant. 
Details of the calculation that produced this figure are provided below.

The dilute regime is characterized by collision rates small compared to the trap frequencies.  Hence in this case, pseudotemperature in the warm, $z$ direction gradually diminishes, while that in the other, cooler directions gradually increases, until the gas equilibrates on the time scale shown.  The hydrodynamic gas, by contrast, behaves like a somewhat compressible fluid;  excitation initially in the $z$ direction is distributed almost immediately into the other directions, and the resulting dynamics is more like the irregular flow to and fro of this liquid about its stationary center of mass. The fluid expands sometimes in the radial direction, sometimes in the axial direction, with irregularly varying amplitudes, reminiscent of waves on an unquiet ocean.
We therefore refer to this form of collective fluid excitation as {\it weltering}. \footnote{ Thus in a tense moment for Ahab aboard a whale boat, Melville relates that the ``... White Whale darted through the weltering ocean.'' ({\it Moby Dick}, Chap. 135.) }

In the dilute gas case, the primary response of the gas is to come to thermal equilibrium, whereby its dynamics is largely summarized in a single, density-normalized equilibration rate, whose inverse defines the ``number of collisions per rethermalization'' \cite{Monroe93_PRL}. For dipolar gases, this quantity can depend on the orientation of the dipoles relative to the excitation axis \cite{Bohn14_PRA, Wang21_PRA}. Vice versa, the complex dynamics of the hydrodynamic fluid requires a more complete theoretical description.

The purpose of this paper is to provide such a description.  We will base full dynamics on a Monte Carlo simulation, to further elaborate the difference between dilute and hydrodynamic regimes.  Further, we will develop a simplified formulation based on a Gaussian {\it ansatz} for the width of a gas, which semi-empirically reproduces the numerics.  Key to this model is the realization that the periphery of a harmonically trapped gas is always dilute \cite{Schafer14_PRA, Bluhm15_PRA}, which necessitates defining an effective volume inside which hydrodynamics is a good idea.  We identify the dependence of this volume on the anisotropy of the trap and of the collision cross section among polarized dipoles. 
Our theory is also presented in a manner that accommodates arbitrary elastic cross sections, 
opening its applicability to a broader variety of ultracold molecular gas experiments with far from threshold collisions \cite{Sadeghpour00_JPB}.

The remainder of this paper is organized as follows: 
In Sec.~\ref{sec:numerical_studies}, we describe the numerical tools adopted to study trapped hydrodynamic gases, and present notable differences from the dilute limit.
We then introduce the equations of motion employed to model a nondegenerate hydrodynamic dipolar gas in Sec.~\ref{sec:hydrodynamic_formulation}, with the assumption of threshold scattering. A variational ansatz is employed in Sec.~\ref{sec:variational_ansatz_method}, to derive effective dynamical equations governing weltering oscillations in a harmonic trap. 
A comparison of our theory to full numerical solutions is presented in Sec.~\ref{sec:hydrodynamic_volume}, from which we purport several considerations about the hydrodynamic extent of gases in traps. 
Finally, conclusion are drawn in Sec.~\ref{sec:conclusions}, along with possible extensions of this current work.

\section{ Numerical method } \label{sec:numerical_studies}

A gas is said to be hydrodynamic when the molecular mean-free path is much smaller than the characteristic length over which fluid flow occurs \cite{Huang63_NY}. The ratio of these scales is given by the Knudsen number ${\rm Kn}$. For a harmonically trapped gas with mean density $\langle n \rangle = \frac{1}{N} \int n^2(\boldsymbol{r}) d^3 r$ and  molecules with total cross section $\sigma_{\rm coll}$, the mean-free path is given by
$L = ( \langle n \rangle \sigma_{\rm coll} )^{-1}$.  
With a given geometric mean frequency $\overline{\omega}$ and temperature $T$, the thermal width of the gas is
 $R_{\rm th} = \sqrt{ k_B T / m \overline{\omega}^2 }$.
 
Alternatively, the Knudsen number can also be written as the ratio of mean trapping frequency over the collision rate $\gamma_{\rm coll} = \langle n \rangle \sigma_{\rm coll} \langle v_{\rm coll} \rangle$, where $\langle v_{\rm coll} \rangle = \sqrt{ 16 k_B T / ( \pi m ) }$ is the mean collision velocity. Explicitly, these relations are summarized as  
\begin{align} \label{eq:Knudsen_number}
    {\rm Kn} 
    &= 
    \frac{ L }{ R_{\rm th} } 
    =
    \frac{ 4 \: \overline{\omega} }{ \pi^{\sfrac{1}{2}} \gamma_{\rm coll} } 
    =
    \frac{ 8 \pi^{\sfrac{3}{2}} k_B T }{ N m \overline{\omega}^2 \sigma_{\rm coll} }.
\end{align}
A trapped gas is said to be hydrodynamic if ${\rm Kn} \ll 1$. 
The relations above provide an approximate mean Knudsen number. In practice, the thermal width can differ in directions with different trap frequencies, while the cross section, for dipolar scattering, can depend on the direction of the collisions axis. Thus the boundary between hydrodynamic and dilute flow can be anisotropic, a topic to be dealt with below.

To compute dynamics in either regime, we utilize the direct simulation Monte Carlo (DSMC) method \cite{Bird70_PF} to obtain numerical solutions to the Boltzmann equation. 
In doing so, these numerical simulations allow for explorations of hydrodynamic phenomena, while later also serving as a benchmark for our semi-empirical theory.

The DSMC implementation we adopt for this work follows very closely that described in Refs.~\cite{Sykes15_PRA, Wang20_PRA}, which study similar systems but in the dilute regime. 
Described briefly, the Boltzmann equation is solved by approximating the phase space distribution with a discrete ensemble of $N$ molecules
\begin{align} \label{eq:phasespace_discretized}
    f(\boldsymbol{r}, \boldsymbol{v}) 
    \approx
    \sum_{ k = 1 }^{ N } \delta^3(\boldsymbol{r} - \boldsymbol{r}_k) \delta^3(\boldsymbol{v} - \boldsymbol{v}_k).
\end{align} 
Most crucial to an accurate hydrodynamic simulation is that collisions are handled adequately. The DSMC does so by constructing a discrete spatial grid within the simulation volume, binning particles into each grid cell based on their positions, then sampling their collisional interactions from a probability distribution derived from the differential cross section \cite{Sykes15_PRA}. 

Choosing a uniform grid that is appropriate for maintaining accuracy and computational efficiency becomes tricky at large collision rates, so we utilize a locally adaptive discretization scheme instead. 
At every numerical time step, the locally adaptive grid is built in two phases. Phase one constructs a master grid, consisting of uniform volume cells that span the simulation volume. 
The resolution of the grid is then refined in phase two, with an octree algorithm \cite{Franklin85_Springer}. The octree algorithm further discretizes the simulation volume by recursively subdividing cells into eight octants, terminating when each cell has at most $N_{\text{cell}}^{\max}$ particles. The parameter $N_{\text{cell}}^{\max}$, is initialized at the start of the simulation, which we optimize for stochastic convergence.  

\section{ Numerical Results }

\begin{table}[ht]
\caption{ \label{tab:system_parameter} 
Table of parameter values utilized in the Monte Carlo simulation for fermionic $^{23}$Na$^{40}$K molecules. Da $= 1.661 \times 10^{-27}$ kg stands for Dalton (atomic mass unit) and ${\rm D} = 3.33564 \times 10^{-30}$ C$\cdot$m is a Debye. }
\begin{ruledtabular}
\begin{tabular}{l c c c}
    \multicolumn{1}{c}{\textrm{Parameter}} & \multicolumn{1}{c}{\textrm{Symbol}} & \multicolumn{1}{c}{\textrm{Value}} & \multicolumn{1}{c}{\textrm{Unit}} \\
    \colrule
    Relative molecular mass & $M_r$ & 63 & Da \\
    Electric dipole moment & $d$ & 0.75 & D \\
    Initial gas temperature & ${\cal T}(0)$ & 700 & nK \\
    Trap frequency geometric mean & $\overline{\omega}$ & $2 \pi 100$ & Hz 
\end{tabular}
\end{ruledtabular}
\end{table}

For our numerical experiments, we envision an ultracold gas of microwave shielded $^{23}$Na$^{40}$K molecules with the parameters in Tab.~\ref{tab:system_parameter}. The initial temperature is chosen such that the gas remains nondegenerate with $T > T_F$ \cite{Butts97_PRA} for all values of ${\rm Kn}$ in consideration, and the trap is assumed cylindrically symmetric with $\omega_x = \omega_y \equiv \omega_{\perp}$ but $\omega_{\perp} \neq \omega_z$.
Key variables of interest to this study will be: 
a) the number of molecules $N$, which affects ${\rm Kn}$ and therefore how hydrodynamic the gas is; b) the trap anisotropy $\lambda = ( \omega_z / \omega_{\perp} )^2$; c) and the dipole orientation $\hat{\boldsymbol{{\cal E}}}$. For the sake of illustration, collision cross sections are described by the analytical formulas for point dipoles given in Ref.~\cite{Bohn14_PRA}, although at sufficient temperature, realistic cross sections may differ from these.
For convenience, we only allow $\hat{\boldsymbol{{\cal E}}}$ to tilt within the $x,z$-plane, allowing us to define a dipole tilt angle $\Theta = \cos^{-1}\hat{\boldsymbol{{\cal E}}} \cdot \hat{\boldsymbol{z}}$, that parameterizes the collisional anisotropy.

\begin{figure}[ht]
    \centering
    \includegraphics[width=0.95\columnwidth]{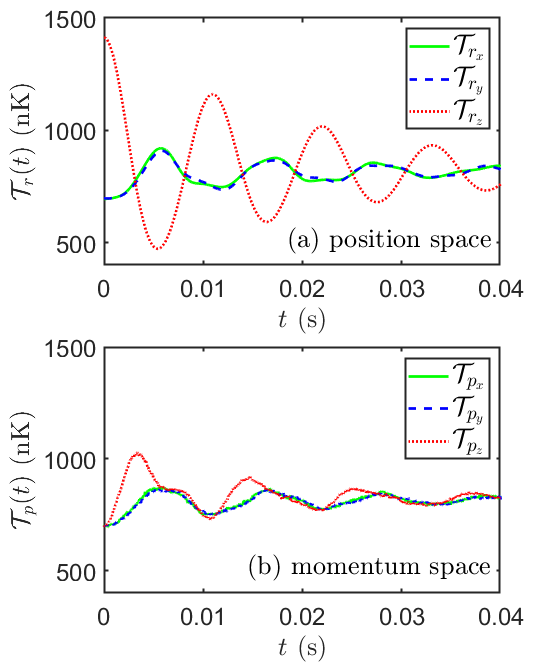}
    \caption{ Plots of the ${\cal T}_{r_i}$ (upper panel a) and ${\cal T}_{v_i}$ (lower panel b) vs time from a cross-dimensional rethermalization experiment, with excitation along $z$. The gas is hydrodynamic with $N = 5 \times 10^5$ (${\rm Kn} \approx 0.04$), $\lambda = 0.2$ and the parameters in Tab.~\ref{tab:system_parameter}. }
    \label{fig:Tr_vs_Tp_HD_Theta90}
\end{figure}

The behavior of the fluid after excitation in the $z$ direction is shown in Fig.~\ref{fig:Tr_vs_Tp_HD_Theta90}.
This is done in a prolate (cigar) trap with $\lambda = 0.2$, containing $N = 5 \times 10^{5}$ molecules, with Knudsen number ${\rm Kn} \approx 0.04$. 
This figure  plots  the separated position and momentum space pseudotemperatures ${\cal T}_{r_i}(t) = m \omega_i^2 \{ r_i^2 \}(t) / k_B$ and ${\cal T}_{v_i}(t) = m \{ v_i^2 \}(t) / k_B$ respectively. The position space time trace shows the clear out-of-phase oscillations between the widths in the radial and axial directions, expected for a weltering fluid. 
The momentum space time trace has oscillations of considerably smaller magnitude than ${\cal T}_{r_i}$, and also shows a phasing in oscillations amongst the different ${\cal T}_{v_i}$ traces. These observations showcase how large collision rates diminish the effect of out-of-equilibrium thermodynamics on the hydrodynamic welter of the gas. 

\begin{figure}[ht]
    \centering
    \includegraphics[width=\columnwidth]{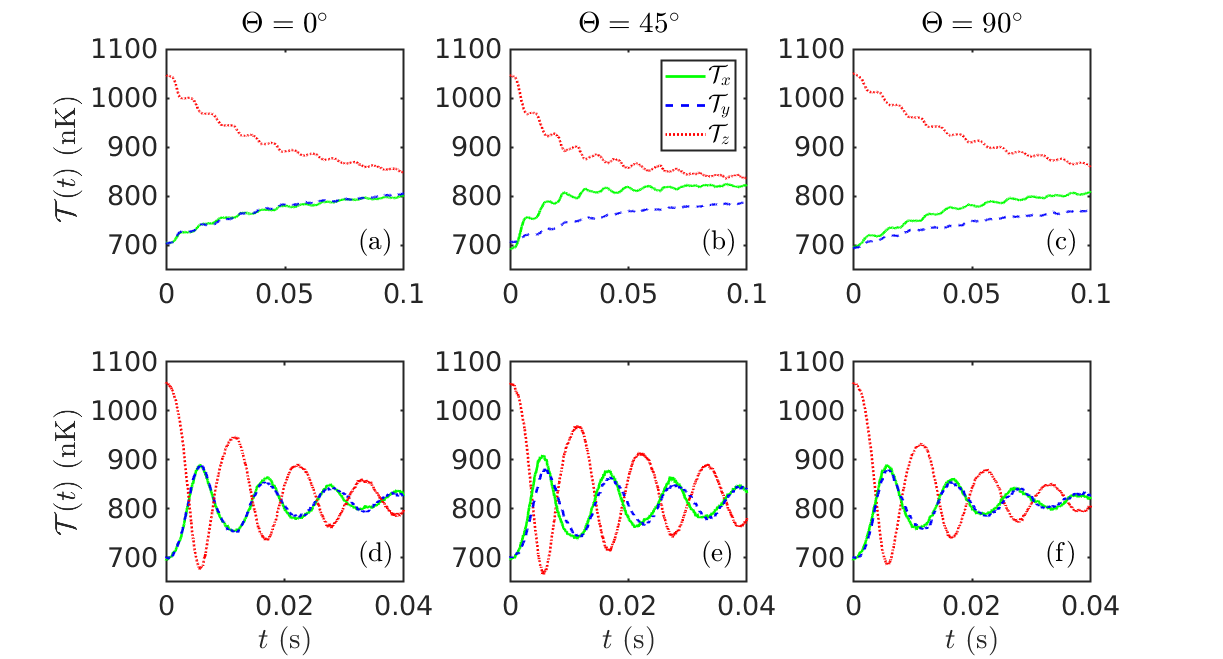}
    \caption{ Pseudotemperature times traces ${\cal T}_x(t)$ (solid green curves), ${\cal T}_y(t)$ (dashed blue curves) and ${\cal T}_z(t)$ (dotted red curves) for 3 values of $\Theta = 0^{\circ}, 45^{\circ}, 90^{\circ}$, in subplots (a, d), (b, e) and (c, f) respectively.
    The 2 rows are differentiated by the number of molecules, with the upper row (subplots a, b, c) having $N = 2 \times 10^3$ (${\rm Kn} \approx 11.10$), while the lower row (subplots d, e, f) has $N = 3 \times 10^5$ (${\rm Kn} \approx 0.07$).
    The experimental parameters are those in Tab.~\ref{tab:system_parameter} with $\lambda = 0.2$.
    Note that the simulation times are different between the upper ($t = 0$ to $0.1$s) and lower ($t = 0$ to $0.04$s) rows. }
    \label{fig:Txyz_vs_Theta_(DSMC)}
\end{figure}

The difference between dilute and hydrodynamic regimes is sharpened by comparing the dependence of dynamics on the tilt angle $\Theta$ of the dipoles. To this end, Fig.~\ref{fig:Txyz_vs_Theta_(DSMC)} plots the three components of pseudotemperature ${\cal T}_{i}$ for the dilute (upper row) and hydrodynamic (lower row) gases, at the 3 different dipole tilt angles $\Theta = 0^{\circ}, 45^{\circ}, 90^{\circ}$.

As anticipated in Fig.~\ref{fig:pseudotemps_DLvsHD_Theta90}, the dilute gas responds to the excitation primarily by melting back to thermal equilibrium while the hydrodynamic gas exhibits radial weltering motion, resulting from oscillating fluid flow toward and away from the trap center.
In Fig.~\ref{fig:Txyz_vs_Theta_(DSMC)} a second dramatic difference appears.  For the dilute gas, with the dipoles tilted away from the axis of trap symmetry ($z$), the rates of warming of the gas in the $x$ and $y$ directions differ, as a consequence of the anisotropic scattering cross section \cite{Bohn14_PRA, Sykes15_PRA, Wang21_PRA}.  By contrast, the   excitations in the $x$ and $y$ directions in the hydrodynamic regime are nearly equal.   In the hydrodynamic regime, relatively rapid collisions scramble memory of the dipole orientation.  Note that a slight difference in $x$ and $y$ motions occurs, due to a residual anisotropy of the viscosity tensor, described in the next section.  Nevertheless, this anisotropy is not a main driving force in the dynamics. It is true, however, that the overall damping rate of the weltering excitations does depend on the dipole tilt angle, as will be elaborated upon in Sec.~\ref{sec:hydrodynamic_volume} below.

\onecolumngrid

\twocolumngrid

\section{ Hydrodynamic formulation \label{sec:hydrodynamic_formulation} }

The Monte Carlo simulation, while accurate, is nevertheless somewhat cumbersome for calculating the response of the gas.  For this reason, in the hydrodynamic regime, it is useful to formulate the fluid's motion directly in terms of hydrodynamics.
When hydrodynamic, a nondegenerate gas behaves as a thermoviscous fluid \cite{Griffin97_PRL, Nikuni98_JLTP, Kavoulakis98_PRA} with thermal conductivity $\kappa_{i j}$, and viscosity $\mu_{i j k \ell}$, which are, in general, coordinate dependent and formulated as rank-2 and rank-4 tensors respectively \cite{deGroot13_DP}. 
The equations of motion of the fluid are \cite{Fetter03_Dover}:
\begin{subequations} \label{eq:continuum_conservation_laws_trapped}
\begin{align} \label{eq:continuity}
    \frac{ \partial \rho }{ \partial t } 
    +
    \sum_j \partial_j \left( { \rho U_j } \right) 
    &= 
    0, \\ \label{eq:Navier_Stokes}
    \frac{\partial}{\partial t} \left( \rho U_i \right) 
    +
    \sum_j \partial_j \left( \rho U_j U_i \right) 
    &= 
    -
    \partial_i 
    \left( n k_B T \right)
    - 
    n \partial_i V(\boldsymbol{r}) \nonumber\\ 
    & +
    \sum_{j, k, \ell} 
    \partial_j 
    \left( \mu_{ i j k \ell } \partial_{\ell} U_k \right), \\ \label{eq:temperature_balance}
    \frac{\partial}{\partial t} (\rho T) 
    +
    \sum_j \partial_j \left( \rho T U_j \right) 
    &= 
    -
    \frac{ 2 }{ 3 }
    \rho T
    \sum_i \partial_i U_i \nonumber\\
    & +
    \frac{ 2 m }{ 3 k_B } 
    \sum_{i, j, k, \ell} 
    ( \partial_j U_i )
    \mu_{ i j k \ell }
    ( \partial_{\ell} U_k )
    \nonumber\\
    & +
    \frac{ 2 m }{ 3 k_B } 
    \sum_{i, j} 
    \partial_i \left( \kappa_{i j} \partial_j T \right).
\end{align}
\end{subequations}
These equations govern the dynamics of the velocity averaged field variables of mass density, flow velocity and temperature:
\begin{subequations}
\begin{align}
    \rho(\boldsymbol{r}, t) 
    &= 
    m n(\boldsymbol{r}, t) = \int d^3 v f(\boldsymbol{r}, \boldsymbol{v}, t) m, \\
    \boldsymbol{U}(\boldsymbol{r}, t) 
    &= 
    \frac{1}{n(\boldsymbol{r}, t)} \int d^3 v f(\boldsymbol{r}, \boldsymbol{v}, t) \boldsymbol{v}, \\
    T(\boldsymbol{r}, t) 
    &= 
    \frac{ 2 }{ 3 n(\boldsymbol{r}, t) k_B } \int d^3 v f(\boldsymbol{r}, \boldsymbol{v}, t) \frac{1}{2} m \boldsymbol{u}^2, \label{eq:local_temperature}
\end{align}
\end{subequations}
where $f(\boldsymbol{r}, \boldsymbol{v}, t)$ denotes the phase space distribution of the molecules and $\boldsymbol{u}(\boldsymbol{r}) = \boldsymbol{v} - \boldsymbol{U}(\boldsymbol{r})$ is the comoving molecular velocity, relative to the frame of fluid flow. 

It is worth pointing out that the local fluid kinetic temperature is related to the flow velocity via
\begin{align} \label{eq:T_vs_U}
    \frac{ 3 }{ 2 }
    n(\boldsymbol{r}, t) k_B T(\boldsymbol{r}, t)
    &=
    \int d^3 v f(\boldsymbol{r}, \boldsymbol{v}, t) \frac{ 1 }{ 2 } m \boldsymbol{v}^2 \nonumber\\
    &\quad -
    \frac{ 1 }{ 2 } \rho \boldsymbol{U}(\boldsymbol{r}, t)^2,
\end{align}
where the integral term is the local kinetic energy density. 
This relation emphasizes a central difference between dilute and hydrodynamic trapped gases: temperature, in the sense of equilibrium thermodynamics, is well defined throughout the entire dynamical evolution when hydrodynamic, but only upon global equilibration when dilute. Such a distinction identifies time-of-flight imaging, common to ultracold gas experiments, as an indirect form of thermometry to hydrodynamic gases, that probes an ensemble averaged sum of both the fluid local temperature and mechanical energy from flow.

In this work, we assume that the transport tensors arise from two-body collisions with elastic differential cross section $d\sigma / d\Omega$, as derived with the first-order Chapman-Enskog method \cite{Chapman90_CUP, Wang22_PRA, Wang22_PRA2}. We shall later see that only viscosity is relevant to this work, so we omit further details of the thermal conductivity. 
At this level of approximation, the anisotropic viscosity tensor for arbitrary $d\sigma / d\Omega$ works out to be density independent, and is given as \cite{Wang22_PRA2, Wang23_PRA} 
\begin{align} \label{eq:viscosity_tensor} 
    \boldsymbol{\mu} 
    &=
    -\frac{ 2 }{ \beta } \left( \frac{ n }{ m \beta } \right)^2 
    \left(
    \int d^3 u \boldsymbol{W}(\boldsymbol{u}) \otimes C[ f_0 \boldsymbol{W} ]
    \right)^{-1},
\end{align}
where $\beta = ( k_B T )^{-1}$ is the usual inverse temperature, 
\begin{align} 
    \boldsymbol{W}
    &=
    \boldsymbol{u} \boldsymbol{u}^T
    -
    \frac{ 1 }{ 3 }
    \boldsymbol{u}^2
    \boldsymbol{I},
\end{align} 
is a rank-2 comoving velocity tensor, and $\boldsymbol{I}$ is the identity matrix. 
The collision integrals
\begin{align} \label{eq:collision_integral}
    C[ f_0 \boldsymbol{W} ]
    &=
    \int d^3 u_1 
    \abs{ \boldsymbol{u} - \boldsymbol{u}_1 }
    f_0(\boldsymbol{u}) f_0(\boldsymbol{u}_1)
    \int d\Omega'
    \frac{ d\sigma }{ d\Omega' }
    \Delta\boldsymbol{W},
\end{align}
with $\Delta\boldsymbol{W} = \boldsymbol{W}' + \boldsymbol{W}'_1 - \boldsymbol{W} - \boldsymbol{W}_1$ and primes denoting post-collision quantities, 
are evaluated with the Maxwell-Boltzmann equilibrium phase space distribution function $f_0(\boldsymbol{u})$ \cite{Reif09_Waveland}. 
The symbol $\otimes$ denotes a dyadic product which takes two tensors of rank $N_1$ and $N_2$, and forms a tensor of rank $N_1 + N_2$ (e.g. $A_{i j} \otimes B_{k \ell} = C_{i j k \ell}$).  
Of interest here is the anisotropic cross section resultant from close-to-threshold scattering 
\footnote{ With the parameters in Tab.~\ref{tab:system_parameter}, the typical collision energies are in fact much larger than the dipole energy, which may place the assumption of threshold scattering on shaky grounds. Nevertheless, we proceed with the threshold approximation in this work. See App.~\ref{app:threshold_consideration} for further discussions. }
between ultracold fermionic polar molecules or dipolar atoms \cite{Matsuda20_Sci, Schindewolf22_Nat, Patscheider21_PRA, Tang15_PRA}. 
At low enough temperatures with electric fields that align the dipoles along $\hat{\boldsymbol{{\cal E}}}$, dipolar scattering is energy independent and permits the viscosity tensor to be computed analytically \cite{Wang23_PRA}. It is this analytic viscosity tensor that we use below.

\subsection{ Viscous damping of a trapped fluid \label{sec:variational_ansatz_method} }

\begin{figure}[ht]
    \centering
        
        
        
    \includegraphics[width=0.85\columnwidth]{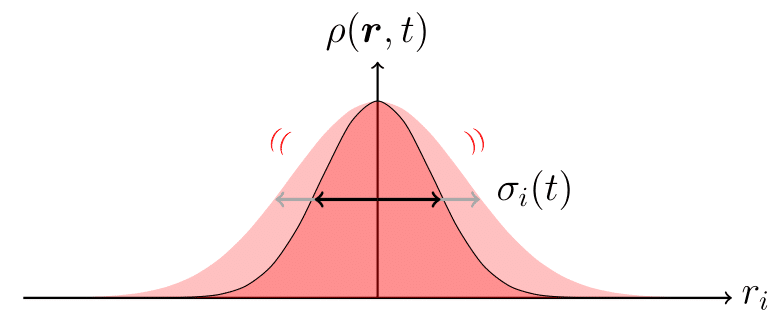}
    \caption{ Cartoon of a density slice  along axis $r_i$, through the Gaussian \textit{ansatz} for $\rho(\boldsymbol{r}, t)$ with time varying widths $\sigma_i(t)$. }
    \label{fig:nonequilibrium_ansatz}
\end{figure}

The fluid equations in (\ref{eq:continuum_conservation_laws_trapped}) are highly nonlinear and, in general, require numerical methods to obtain solutions.
For our purposes, we instead adopt a variational ansatz approach to solving these partial differential equations \footnote{
Our method has similarities to the scaling ansatz employed in Ref.~\cite{Kagan97_PRA, Pedri03_PRA, Wachtler17_PRA}, but formulated slightly differently and with the inclusion of transport tensors. 
}.
External confinement from a harmonic potential results in the equilibrium (denoted by subscript $0$) density distribution following
\begin{align} \label{eq:equilibrium_density_distribution}
    \rho_0(\boldsymbol{r})
    &=
    \frac{ m N }{ Z }
    \exp( -\frac{ V(\boldsymbol{r}) }{ k_B T_0 } ),
\end{align}
where $Z = \int d^3 r {\rm e}^{ -\frac{ V(\boldsymbol{r}) }{ k_B T_0 } }$ gives the appropriate normalization and $N$ is the number of molecules.
If we were then only to consider collective oscillations and damping from long wavelength excitations that do not induce center-of-mass sloshing, Eq.~(\ref{eq:equilibrium_density_distribution}), motivates a Gaussian variational ansatz for the local density:
\begin{align} \label{eq:Gaussian_density_ansatz}
    \rho(\boldsymbol{r}, t)
    =
    m N
    \prod_{i=1}^3
    \frac{ 1 }{ \sqrt{ 2 \pi \sigma_i^2(t) } }
    \exp\left(
    -\frac{ r_i^2 }{ 2 \sigma_i^2(t) }
    \right),
\end{align}
where $\sigma_i(t)$ is the distribution widths along each axis $i$ that we allow to vary in time (depicted in Fig.~\ref{fig:nonequilibrium_ansatz}).

\begin{figure}[ht]
    \centering
    \includegraphics[width=\columnwidth]{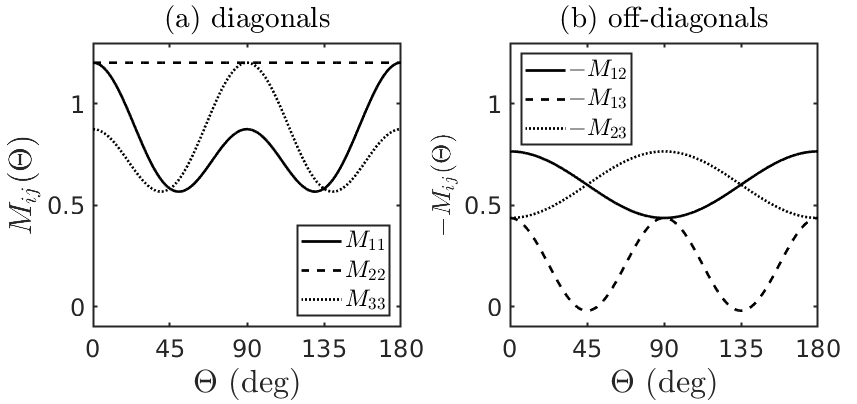}
    \caption{ $M_{i j}$ matrix elements as a function of $\Theta$. The diagonal elements are plotted on the left in subplot (a), whereas the negated (multiplied by a minus sign) off-diagonal elements plotted on the right in subplot (b). }
    \label{fig:Mij_elements}
\end{figure}

Plugging the ansatz of Eq.~(\ref{eq:Gaussian_density_ansatz}) into the continuity equation (\ref{eq:continuity}) gives
\begin{align}
    \sum_{i = 1}^3
    \bigg[
    \partial_i U_i(\boldsymbol{r})
    &-
    U_i (\boldsymbol{r}) \left( \frac{ r_i }{ \sigma_i^2(t) } \right) \nonumber\\
    &\quad+
    \left( \frac{ r_i^2 }{ \sigma_i^2(t) } - 1 \right)
    \frac{  \dot{\sigma}_i(t) }{ \sigma_i(t) }
    \bigg]
    =
    0,
\end{align}
which admits the velocity field solution
\begin{align} \label{eq:flow_velocity_ansatz}
    U_i(\boldsymbol{r})
    =
    \left(
    \frac{ \dot{\sigma}_i(t) }{ \sigma_i(t) }
    \right) r_i.
\end{align}
Thus, as expected, the fluid flow vanishes in the trap's center for the excitations we consider.
These functional forms for $\rho$ and $\boldsymbol{U}$ then render the Navier-Stokes equation (\ref{eq:Navier_Stokes}), of the form
\begin{align} \label{eq:sigma_with_T} 
    \ddot{\sigma}_i(t) 
    +
    \omega_i^2 \sigma_i(t)
    &= 
    \frac{ k_B }{ m }
    \left( \frac{ 1 }{ \sigma_i(t) }
    -
    \frac{ \sigma_i(t) }{ r_i }
    \partial_i
    \right)
    T(\boldsymbol{r}, t) \nonumber\\[2.5pt]
    &\quad +
    \sigma_i 
    \sum_{j, k, \ell} 
    \frac{ \partial_j \mu_{i j k \ell}(T) }{ r_i \rho(\boldsymbol{r}) } 
    \delta_{k, \ell}
    \frac{ \dot{\sigma}_k }{ \sigma_{\ell} },
\end{align}
which bears no dependence on the thermal conductivity. 
Since $\sigma_i(t)$ does not depend on spatial coordinates, consistency requires that we take a spatial average to suppress local fluctuations of the temperature field in Eq.~(\ref{eq:sigma_with_T}). This average is taken by multiplying Eq.~(\ref{eq:sigma_with_T}) and the temperature balance equation (\ref{eq:temperature_balance}),
by $n(\boldsymbol{r}, t)$, then integrating over $d^3 r$. App.~\ref{app:spatial_averaging} gives further details of the spatial averaging procedure, which results in
\begin{widetext}
\begin{align} \label{eq:widths_ODE_Tindependent}
    \ddot{\sigma}_i(t) 
    & +
    \omega_i^2 \sigma_i(t)
    +
    \frac{ 1 }{ 3 \sigma_i(t) }
    \sum_j
    \left[
    \omega_j^2 \sigma_j^2(t) 
    +
    \dot{\sigma}_j^2(t)
    \right]
    -
    \frac{ 2 k_B T_0 }{ m \sigma_i(t) } 
    \approx  
    - \frac{ 2 }{ 5 }
    \frac{ {\cal V}_{\rm hy} }{ N m } 
    \sum_{j} 
    \frac{ \mu_{i i j j}( T(t) ) }{ \sigma_i(t) }
    \frac{ \dot{\sigma}_{j}(t) }{ \sigma_{j}(t) }.
\end{align}
The relevant viscosity matrix elements can be recast in terms of a unit-free matrix
\begin{align} \label{eq:viscosity_matrix}
    M_{i j}(\Theta)
    &\equiv
    \frac{ \mu_{i i j j}(T; \Theta) }{ \mu_0(T) } \nonumber\\
    &= 
    \frac{ 1 }{ 512 }
    \begin{pmatrix}
         117 \cos(4\Theta) + 84 \cos(2\Theta) + 415  & 
        -28 ( 3 \cos(2\Theta) + 11 ) & 
        -( 117 \cos(4\Theta) + 107 ) \\[2.5pt]
        -28 ( 3 \cos(2\Theta) + 11 ) & 
        616 & 
        28 ( 3 \cos(2\Theta) - 11 ) \\[2.5pt]
        -( 117 \cos(4\Theta) + 107 ) & 
        28 ( 3 \cos(2\Theta) - 11 ) &
        117 \cos(4\Theta) - 84 \cos(2\Theta) + 415
    \end{pmatrix},
\end{align}
\end{widetext} 
as is taken from Ref.~\cite{Wang23_PRA}, where the isotropic unit-full viscosity coefficient is given by \cite{Chapman90_CUP} 
\begin{align} \label{eq:CE_viscosity_coefficient} 
    \mu_0(T)
    &=  
    \frac{ 5 }{ 16 a_d^2 } \sqrt{ \frac{ m k_B T }{ \pi } }.
\end{align}
With the parameters in Tab.~\ref{tab:system_parameter}, the isotropic viscosity has a value of $\mu_0 \approx 2.5 \times 10^{-15}$ Pa$\cdot$s, which is around $10^{10}$ times less than air at room temperature and pressure \cite{Kadoya85_JPCRD}.
The $M_{i j}(\Theta)$ matrix elements are plotted in Fig.~\ref{fig:Mij_elements}, with components coupled to the $x$ and $z$ axes showcasing a significant variation with $\Theta$. We see in Fig.~\ref{fig:Mij_elements} that the magnitude of off-diagonal matrix elements $M_{1 3} = M_{x z}$
and $M_{2 3} = M_{y z}$ become maximally separated around $\Theta \approx 45^{\circ}$, explaining the slight separation of ${\cal T}_x(t)$ and ${\cal T}_y(t)$ in Fig.~\ref{fig:Txyz_vs_Theta_(DSMC)}, otherwise negligible when $\Theta = 0^{\circ}, 90^{\circ}$.  

Eq.~(\ref{eq:widths_ODE_Tindependent}) above treats the temperature field appearing in $\mu_{i j k \ell}(T)$ to be spatially uniform over the region where the gas is hydrodynamic. Such an approximation follows from the form of collective oscillations implied by the density (\ref{eq:Gaussian_density_ansatz}) and flow velocity fields (\ref{eq:flow_velocity_ansatz}) in an initially isothermal gas, disallowing a spatial temperature variation on the order of the gas spatial widths \cite{Kavoulakis98_PRA, Schafer14_PRA}. Hence, temperature as appears in the viscosity is simply treated as $T \approx T(t)$. 
In doing so, we were required to define an effective hydrodynamic volume ${\cal V}_{\rm hy} = \int d^3 r$ 
\footnote{Refs.~\cite{Kavoulakis98_PRA, Schafer07_PRA} provides another estimate for the effective hydrodynamic volume which, however, we find systematically underestimates the results from our numerical Monte Carlo simulations. }. 
Proper identification of this volume, including its dependence on aspect ratio, density, and dipole tilt, is essential to the performance of the model, and is our main undertaking here.  We define this volume to be the spheroidal volume bounded by the outer classical turning radius of the trap, multiplied by an empirical factor $\eta$. 
The outer turning radius is obtained by equating $E_{\rm total} = V( R_{\rm HD}, \theta, \phi )$, to give (see App.~\ref{app:spatial_averaging})
\begin{align}
    R_{\rm HD}^2( \theta )
    &=
    \frac{ 6 k_B T(t) }{ m \omega_{\perp}^2 }
    \left[ \sin^2\theta + \lambda \cos^2\theta \right]^{-1},
\end{align}
where $\lambda = (\omega_z / \omega_{\perp})^2$ quantifies the trapping anisotropy. 
The effective hydrodynamic volume is then computed as 
\begin{align} \label{eq:HD_volume}
    {\cal V}_{\rm hy}( \lambda, {\rm Kn} )
    &= 
    \frac{ \eta( \lambda, {\rm Kn} ) }{ 3 } 
    \int R_{\rm HD}^3( \Omega ) d\Omega \nonumber\\
    &=
    \frac{ 4 \pi }{ 3 }
    \left(
    \frac{ 6 k_B T(t) }{ m \omega_{\perp}^2 }
    \right)^{3/2}
    \frac{ \eta( \lambda, {\rm Kn} ) }{ \sqrt{ \lambda } }.
\end{align}
As written, we have assumed that $\eta$ could depend on the trapping geometry through $\lambda$ and on the Knudsen number, which in turn, also implicitly depends on $N$ and the dipole angle $\Theta$. These dependencies are addressed later in the paper.  
Such generality allows $\eta$ to act as a coarse-graining parameter which accounts for all non-hydrodynamic effects excluded from our current theoretical treatment. 
Additionally, Eq.~(\ref{eq:CE_viscosity_coefficient}) implies the temperature dependence of viscosity goes as $\mu_{i i j j}(T) \propto \sqrt{ T }$, for which we will simply approximate as $T \approx T_0$ for all times 
\footnote{
Incorporating time-dependence in the temperature $T(t)$, requires more sophisticated treatments such as second-order hydrodynamics \cite{Lewis17_IOP}. We leave this task to a subsequent study.
}.

For the relevance of time-of-flight imaging, we point out that the momentum space temperature, which differs from the local temperature of Eq.~(\ref{eq:local_temperature}), can also be obtained from solutions to Eq.~(\ref{eq:widths_ODE_Tindependent}) via the relation
\begin{align} \label{eq:momentum_temperature}
    k_B T_{p}(t)
    &=
    \frac{ 1 }{ 3 N } \int d^3 r d^3 v f(\boldsymbol{r}, \boldsymbol{v}, t) m \boldsymbol{v}^2 \nonumber\\
    &= 
    2 k_B T_0
    -
    \frac{ 1 }{ 3 } \sum_i m \omega_i^2 \sigma_i^2(t),
\end{align}
as follows from Eqs.~(\ref{eq:T_vs_U}), (\ref{eq:flow_velocity_ansatz}) and (\ref{eq:total_energy_relation}).

\subsection{ Linear analysis \label{subsec:linear_EOM} }

Some proceeding discussions on collective dynamics are made more accessible in the language of normal modes, motivating a linear analysis of Eq.~(\ref{eq:widths_ODE_Tindependent}). 
If only taken perturbatively out-of-equilibrium, we can consider small deviations away from the equilibrium widths by writing $\sigma_i(t) = \sigma_{0,i} + \delta\sigma_i(t)$. Then expanding to first-order in $\delta\sigma_i(t)$, Eq.~(\ref{eq:widths_ODE_Tindependent}) becomes
\begin{align} \label{eq:linearized_widths_EOM}
    \ddot{\delta\sigma}_i(t)
    +
    2 \sum_j 
    \Gamma_{i j}
    \dot{\delta\sigma}_{j}(t)
    +
    \sum_j
    O_{i j}
    \delta\sigma_j(t)
    &\approx
    0,
\end{align}
with squared-frequency and damping matrices 
\begin{subequations}
\begin{align}
    O_{i j}
    &=
    2 \omega_i^2 \delta_{i,j} 
    +
    \frac{ 2 }{ 3 } 
    \omega_i \omega_j, \\
    \Gamma_{ i j }
    &= 
    \frac{ \mu_0 {\cal V}_{\rm hy} }{ 5 N k_B T_0 }
    \omega_i
    M_{i j}(\Theta)
    \omega_{j}. 
\end{align}
\end{subequations} 
The matrices above encode the anisotropies from both the trap and anisotropic collisions. 
A factor 2 multiplies $\boldsymbol{\Gamma}$ in Eq.~(\ref{eq:linearized_widths_EOM}) as is convention in damped harmonic oscillators. 
With $\boldsymbol{\Gamma}$ multiplying the first-order time derivative terms $\dot{\delta\sigma}_i$, it is made clear that damping of weltering oscillations results from the trap frequency weighted viscosities within the hydrodynamic volume.   

Diagonalizing the squared-frequency matrix $\boldsymbol{O}$ gives the eigenvalues
\begin{subequations} \label{eq:eigenfrequencies}
\begin{align} 
    \omega_0^2 
    &=
    2 \omega_{\perp}^2, \\
    \omega^2_{\pm} 
    &= 
    \frac{ 1 }{ 3 } 
    \left( 4 \lambda + 5 \pm \sqrt{16 \lambda^2 - 32 \lambda + 25} \right)
    \omega_{\perp}^2,
\end{align}
\end{subequations}
which are exactly those obtained for inviscid Euler flow in Refs.~\cite{Griffin97_PRL, Kavoulakis98_PRA}, and correspond to the respective eigenmodes (up to arbitrary normalization)
\begin{subequations} \label{eq:eigenmodes}
\begin{align}
    \boldsymbol{o}_0
    &=
    \begin{pmatrix}
        1 \\ -1 \\0
    \end{pmatrix}, \\
    \boldsymbol{o}_{\pm}
    &=
    \begin{pmatrix}
        5 - 4 \lambda \pm \sqrt{ 25 + 16 \lambda ( \lambda - 2 ) } \\ 
        5 - 4 \lambda \pm \sqrt{ 25 + 16 \lambda ( \lambda - 2 ) } \\ 
        4 \sqrt{\lambda }
    \end{pmatrix}.
\end{align}
\end{subequations}
The eigenmode $\boldsymbol{o}_0$ is a strictly radial quadrupole mode, while $\boldsymbol{o}_{-}$ and $\boldsymbol{o}_{+}$ are 3-dimensional quadrupole and breathing modes respectively.

Similarly, $\boldsymbol{\Gamma}$ results in two nontrivial eigenvalues $\gamma_{\pm}$, that constitute the principle damping rates of the system. Although it is tempting to assign one of these principle rates as the overall relaxation rate, the eigenmodes associated to each $\gamma_{\pm}$, are in general, not the eigenmodes of $\boldsymbol{O}$. Consequently, coupling between the eigenmodes of $\boldsymbol{\Gamma}$ is inevitable during dynamical evolution, enforcing that accurate relaxation trajectories are best obtained from full solutions to Eq.~(\ref{eq:linearized_widths_EOM}).

\subsection{ The hydrodynamic volume \label{sec:hydrodynamic_volume} }

Returning to the main argument, Eq.~(\ref{eq:widths_ODE_Tindependent}) is expected to be a reasonable representation of dynamics, provided the shape of the gas remains nearly Gaussian.  To employ these equations, we must establish the value of the effective hydrodynamic volume.  A first guess at this volume is given in Eq.~(\ref{eq:HD_volume}), which left available a free parameter $\eta$, that may depend on $\lambda$ and ${\rm Kn}$. As noted in Sec.~\ref{sec:variational_ansatz_method}, ${\rm Kn}$ is implicitly dependent on $N$ and $\Theta$, which are taken as the relevant independent variables for this study.

To extract $\eta$, we perform multiple DSMC runs while varying $\lambda$, $N$ and $\Theta$, which provides us time traces of $T_p(t)$ (\ref{eq:momentum_temperature}) for each combination of parameter values. We then fit $T_p(t)$ as computed from our theory (\ref{eq:widths_ODE_Tindependent}) to those from the DSMC simulations while floating $\eta$, 
such that it minimizes the relative root-mean-squared error  
\begin{align} \label{eq:rms_error}
    \varepsilon( \eta )
    =
    \sqrt{
    \sum_t
    \left(
    \frac{ T_p^{\rm DMSC}(t) - T_p^{\rm theory}(t; \eta) }{ T_p^{\rm DMSC}(t) }
    \right)^2
    }.
\end{align} 
In these numerical experiments, we tune the trap anisotropy in a manner that does not the affect ${\rm Kn}$, by setting $\omega_{\perp} = \overline{\omega} / \lambda^{1/6}$ and $\omega_{z} = \overline{\omega} \lambda^{1/3}$. This construction ensures that $\overline{\omega}$, and therefore ${\rm Kn}$, both remain independent of $\lambda$.  
The dipoles are taken to point along $\hat{\boldsymbol{x}}$ for the data shown.  Dependence on dipole orientation will be included below.

\begin{figure}[ht]
    \centering
    \includegraphics[width=\columnwidth]{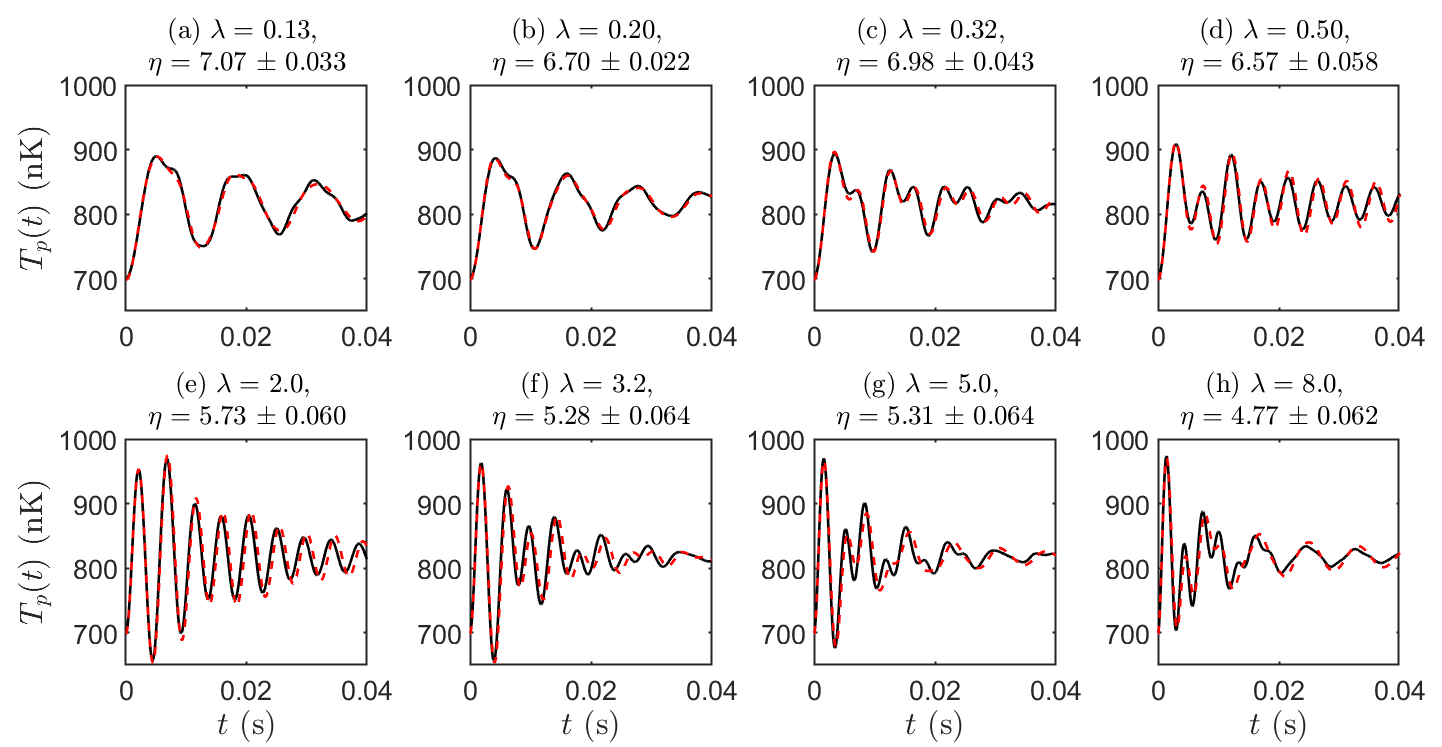}
    \caption{ Comparison of the momentum space temperature $T_p$ (\ref{eq:momentum_temperature}) vs time $t$, obtained from DSMC simulations (black solid curves) and our theory (red dashed curves) with $N = 5 \times 10^5$ (${\rm Kn} \approx 0.04$) and parameters in Tab.~\ref{tab:system_parameter}. The subplots (a) to (h) correspond to various values of trapping anisotropy with $\lambda = 0.13$ to $8.0$ as labeled in the subplot headers.
    The fitted values of $\eta$ are also provided in the subplot headers with their fitting standard uncertainties. }
    \label{fig:TvsLambda_DSMCvsTheory_N5e5}
\end{figure}

\begin{figure}[ht]
    \centering
    \includegraphics[width=0.95\columnwidth]{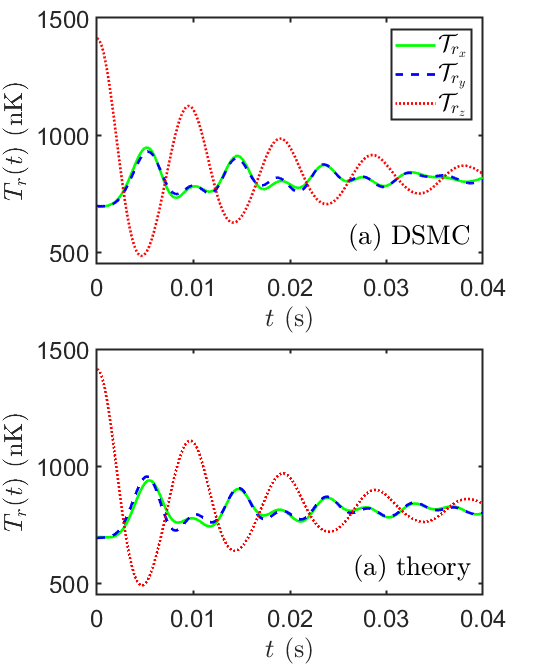}
    \caption{ Comparison of the position space pseudotemperatures ${\cal T}_r$ vs time $t$, obtained from DSMC simulations (upper subplot a) and our theory (lower subplot b) with the parameters in Tab.~\ref{tab:system_parameter}, $N = 5 \times 10^5$ (${\rm Kn} \approx 0.04$) and $\lambda = 0.32$.  }
    \label{fig:Tr_DSMCvsTheory_cigar_Theta90}
\end{figure}

Results of several such  fits are shown in Fig.~\ref{fig:TvsLambda_DSMCvsTheory_N5e5}, which compares the $T_p$ time traces for a series of cross-dimensional rethermalization experiments with $N = 5 \times 10^5$ (${\rm Kn} \approx 0.04$) over a range of $\lambda = 0.13$ to $8.0$, as obtained from DSMC simulations (solid black curves) and our fitted theory (dashed red curves). Noticeably, there is a clear beating of various modes with different frequencies which our theory is able to describe, showing favorable agreement in both the amplitude and phase of oscillations. 
A representative comparison plot of ${\cal T}_r(t)$ as obtained from DSMC and Eq.~(\ref{eq:widths_ODE_Tindependent}) is also provided in Fig.~\ref{fig:Tr_DSMCvsTheory_cigar_Theta90}, with $N = 5 \times 10^5$ (${\rm Kn} \approx 0.04$) and $\lambda = 0.32$. Good agreement is seen in all ${\cal T}_{r_i}(t)$ time traces as well. 
We note that temperature time traces tend to show better agreement to the DSMC ones for excitation along the long axis of a prolate trap, even for larger Knudsen numbers (${\rm Kn} \approx 0.1$). So, we stick to this excitation geometry for a more focused study.

\begin{figure}[ht]
    \centering
    \includegraphics[width=\columnwidth]{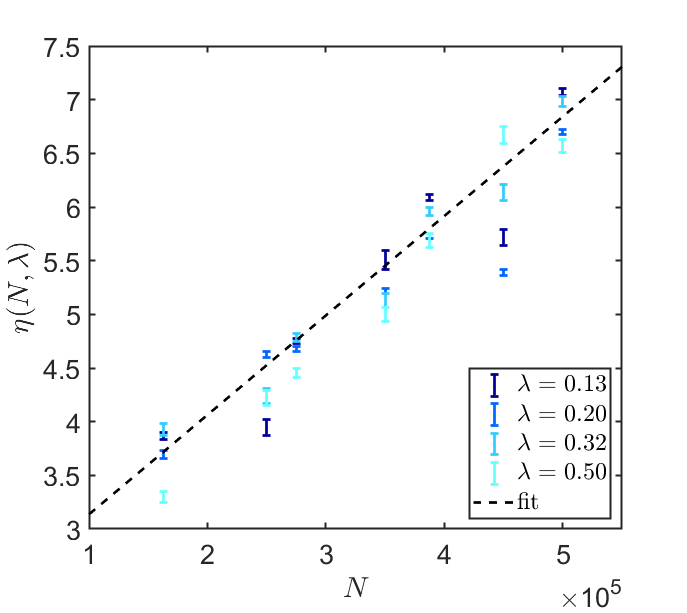}
    \caption{ Plot of $\eta$ vs $N$ for various values of $\lambda = 0.13, 0.20, 0.32, 0.50$, all of which are prolate (cigar) geometries. Also plotted is a linear function ansatz in Eq.~(\ref{eq:eta_ansatz}) (gray dashed line), for comparison with data from DSMC simulations (blue data). Error bars on the DSMC data points denote standard fit uncertainties. }
    \label{fig:ETAvsN_cigar}
\end{figure}

For a given orientation of the dipoles, it may be expected that $\eta$ depends on both the trap aspect ratio $\lambda$ and the number of molecules $N$.  Increasing $N$, {\it ceteris paribus}, evidently increases the density and hence likely the hydrodynamic volume.  As for aspect ratio, a tentative $\lambda$ dependence of ${\cal V}_{\rm hy}$ is already taken into account by (\ref{eq:HD_volume}), whereby the scaling parameter $\eta$ may depend only weakly on $\lambda$. 
This hypothesis is supported by the numerics as shown in Fig.~\ref{fig:ETAvsN_cigar}, where we find that $\eta$ is linearly dependent on $N$, but largely independent of $\lambda$ for the range of these parameters we explore.

Finally, for a given $\lambda$ and $N$, it remains to resolve the dependence of $\eta$ on the dipole orientation $\hat{\boldsymbol{{\cal E}}}$. 
In this context, recall that the dilute and hydrodynamic regimes are distinguished by the Knudsen number, which is inversely proportional to the collision cross section, Eq.~(\ref{eq:Knudsen_number}).   
We saw in Sec.~\ref{sec:variational_ansatz_method}, that this cross section results in anisotropic viscosities, that work to bring local thermodynamic fluctuations back to equilibrium. 
Having accounted for this aspect of differential scattering, we posit that $\eta$ should only depend on the post-collision averaged cross section $\sigma_{\rm coll} = \int d\Omega' \frac{ d\sigma }{ d\Omega' }$, which still preserves an incoming-collision angle dependence \cite{Bohn14_PRA}. As to how so, we present the following argument. 
Prolate traps have a weak trapping axis $z$, along which the gas has a larger thermal width. As a result, the mean-free path along that axis is relatively smaller compared to the sample size, and consequently more hydrodynamic.
Collisions that occur with relative momentum directed along the long axis, are then most able to keep molecules behaving collectively as hydrodynamic.  
The bulk total cross section is, therefore, most simply taken as 
\begin{align} \label{eq:postcollision_averaged_cross_section}
    \sigma_{\rm coll}
    =
    a_d^2 \frac{ \pi }{ 3 }
    \big[
    3 
    +
    18 \cos^2 ( \hat{\boldsymbol{{\cal E}}} \cdot \hat{\boldsymbol{e}}_{\rm hy} )
    -
    13 \cos^4 ( \hat{\boldsymbol{{\cal E}}} \cdot \hat{\boldsymbol{e}}_{\rm hy} )
    \big],
\end{align}
where $\hat{\boldsymbol{e}}_{\rm hy} = \hat{\boldsymbol{z}}$ denotes the most hydrodynamic axis (weakest trap frequency), so that $\hat{\boldsymbol{{\cal E}}} \cdot \hat{\boldsymbol{e}}_{\rm hy} = \Theta$.

\onecolumngrid

\twocolumngrid

\begin{figure}[ht]
    \centering
    \includegraphics[width=\columnwidth]{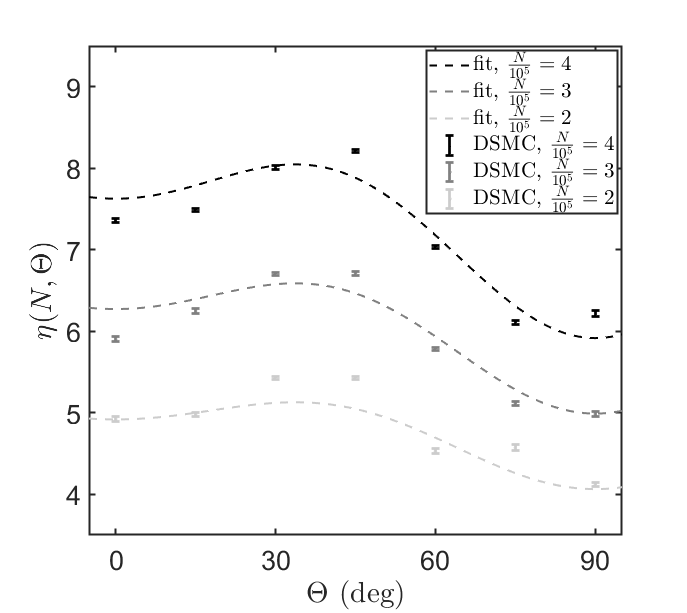}
    \caption{ Plot of $\eta$ vs $\Theta$ from a cross-dimensional rethermalization experiment. The data points (points with error bars) are obtained from DSMC simulations, which is compared to the fitting function (dashed curves) in Eq.~(\ref{eq:eta_ansatz}). The data is obtained with the parameters in Tab.~\ref{tab:system_parameter} and $\lambda = 0.2$, for 3 values of $N = 4 \times 10^5$ (black data, ${\rm Kn} \approx 0.06$), $N = 3 \times 10^5$ (gray data, ${\rm Kn} \approx 0.07$) and $N = 2 \times 10^5$ (light gray data, ${\rm Kn} \approx 0.11$) }
    \label{fig:ETAvsTheta_cigar}
\end{figure}

We indeed find that $\eta$ follows a $\Theta$ dependence very similar to that of Eq.~(\ref{eq:postcollision_averaged_cross_section}), when comparing $\eta$ as obtained from DSMC experiments, to a fitting function of the form $( \sigma_{\rm coll} / \overline{\sigma}_{\rm coll} ) \alpha + \beta$ in Fig.~\ref{fig:ETAvsTheta_cigar}, where $\overline{\sigma}_{\rm coll} = \int \sigma_{\rm coll}(\hat{\boldsymbol{e}}_{\rm hy}) d\hat{\boldsymbol{e}}_{\rm hy} = { 32 \pi a_d^2 / 15 }$ is the angular averaged total cross section.
The observations above motivate the functional form 
\begin{align} \label{eq:eta_ansatz}
    \eta 
    \approx
    a 
    +
    b 
    \left( \frac{ N }{ 10^5 } \right)
    \left[
    1
    +
    c \left( \frac{ \sigma_{\rm coll} }{ \overline{\sigma}_{\rm coll} } \right)
    \right], 
\end{align}
for some constants $a, b$ and $c$, which we determine from fits to be $a \approx 2.21 \pm 0.017$, $b \approx 0.67 \pm 0.020$ and $c \approx 0.26 \pm 0.015$. 
See App.~\ref{app:HD_volume_ansatz} for further details.  
Our functional guess for the hydrodynamic volume is therefore
\begin{align} \label{eq:hydrodynamic_volume_functional_form}
    & {\cal V}_{\rm hy}( \lambda, N, \Theta )
    \approx
    \frac{ 4 \pi }{ 3 }
    \left(
    \frac{ 6 k_B T_0 }{ m \omega_{\perp}^2 }
    \right)^{3/2} \\
    &\quad\quad\quad \times 
    \frac{ 1 }{ \sqrt{ \lambda } }
    \left[
    2.21
    +  
    0.67
    \left( 
    1 
    +
    0.26
    \frac{ \sigma_{\rm coll}(\Theta) }{ \overline{\sigma}_{\rm coll} } 
    \right)
    \frac{ N }{ 10^5 }
    \right]. \nonumber
\end{align}
This quasi-empirical formula constitutes the main result of the present paper. Using this parametrization, the equations of motion  (\ref{eq:widths_ODE_Tindependent}) can be used to reliably determine the evolution of a hydrodynamic dipolar Fermi gas in a prolate trap, subject to excitation along the long axis of the trap.

\section{ Discussions and conclusions \label{sec:conclusions} }

A trapped gas transitions to one that is hydrodynamic when the molecular mean-free path is far exceeded by the extent of its thermal cloud. 
Collisional thermalization is then a local and rapid process, for which collective dynamics becomes likened to that of a fluid. 
In this work, we have studied the damping and oscillations of hydrodynamic welter in harmonically confined dipolar gases, with cross-dimensional rethermalization experiments.

Unlike its dilute counterpart, a hydrodynamic dipolar gas has its distribution width (second moment) dynamics closely follow the symmetries imposed by the confining potential. This adherence to the extrinsic trap symmetry arises from a high frequency of collisions, suppressing the intrinsic dipolar properties from manifesting on macroscopic scales.  
But since local thermal equilibration is not truly instantaneous, dipolar collisions still result in anisotropic viscous shearing between fluid layers, damping the macroscopic fluid welter.  
We have constructed a model to describe such damped weltering dynamics, presented in Eq.~(\ref{eq:widths_ODE_Tindependent}). Embedded in this model is a semi-empirical quantity ${\cal V}_{\rm hy}$, which quantifies the hydrodynamic extent of the trapped gas and its consequence to damping. Through use of numerical experiments, we obtain a functional form for ${\cal V}_{\rm hy}$ in Eq.~(\ref{eq:hydrodynamic_volume_functional_form}), expected to work in the range of $\lambda$, $N$ and $\Theta$ explored here.

Larger Knudsen numbers and trap anisotropies will increase the dilute fraction, requiring more nuanced treatments of the non-hydrodynamic regions.
Moreover, the approximation made in Sec.~\ref{sec:hydrodynamic_formulation} of threshold dipolar scattering, may not be adequate in hydrodynamic samples of polar molecular gases. Threshold scattering requires that the collision energies relative to the dipole energy are sufficiently low \cite{Bohn09_NJP}, but there be high enough collision rates to remain hydrodynamic, as is detailed in App.~\ref{app:threshold_consideration}. This raises issues for Bose gases within the presented formalism, since lowering the temperature to achieve threshold scattering would result in a significant condensate fraction. On the other hand, Fermi gases below $T_F$ still have collective excitations well described by classical kinetic theories, if Pauli blocking effects are included \cite{Nikuni98_JLTP}. 
Lastly, dipolar mean-field effects have been ignored, thermal energies being much larger than the average dipolar mean-field energy per particle \cite{Lahaye09_IOP}. 
All these considerations, albeit important to current molecular ultracold experiments, are not within the current scope of this work and will be considered in future investigations.

\begin{acknowledgments}

The authors would like to thank X. Y. Luo, A. Schindewolf and X. Y. Chen for insightful and motivating discussions on ultracold molecular trapped gases in the hydrodynamic regime.
This work is supported by the National Science Foundation under Grant Number PHY2110327. 

\end{acknowledgments}

\appendix

\section{ Averaging out spatial coordinates \label{app:spatial_averaging} }

To obtain the spatially averaged equations of motion in Sec.~\ref{sec:variational_ansatz_method}, we start by defining a notation for spatially averaged quantities:
\begin{align} \label{eq:spatial_average}
    \langle \ldots \rangle 
    =
    \frac{ 1 }{ N }
    \int 
    n(\boldsymbol{r}, t)
    \left( \ldots \right)
    d^3 r.
\end{align}
This renders the density averaged equation for $\sigma_i(t)$ as
\begin{align} \label{eq:averaged_sigma_diffEQ}
    \frac{ \langle r_i^2 T \rangle }{ \sigma_i^2(t) }
    -
    \langle r_i \partial_i T \rangle 
    &= 
    \frac{ m }{ k_B }
    \left(
    \frac{ \ddot{\sigma}_i(t) }{ \sigma_i(t) }
    +
    \omega_i^2
    \right) 
    \langle r_i^2 \rangle \nonumber\\ 
    &\quad -
    \sum_{j, k, \ell} 
    \frac{ \dot{\sigma}_k }{ \sigma_{\ell} }
    \delta_{k, \ell}
    \int \frac{ d^3 r }{ N k_B }
    r_i \partial_j \mu_{i j k \ell}(T) \nonumber\\ 
    &=
    \frac{ m }{ k_B }
    \left(
    \frac{ \ddot{\sigma}_i(t) }{ \sigma_i(t) } 
    +
    \omega_i^2 
    \right)
    \sigma_i^2(t) \\ 
    &\quad -
    \sum_{j, k, \ell} 
    \frac{ \dot{\sigma}_k }{ \sigma_{\ell} }
    \int \frac{ d^3 r }{ N k_B }
    r_i \partial_j \mu_{i j k \ell}(T)
    \delta_{k, \ell}. \nonumber
\end{align}
As for the temperature balance equation:
\begin{align}
    & \frac{ \partial T(\boldsymbol{r}, t) }{ \partial t }
    + 
    \sum_i U_i \partial_i T(\boldsymbol{r}, t)
    +
    \frac{ 2 }{ 3 } 
    \sum_i 
    \partial_i U_i
    T(\boldsymbol{r}, t) \nonumber\\
    &\quad\quad\quad =
    \frac{ 2 }{ 3 n(\boldsymbol{r}, t) k_B }
    \sum_{i, j, k, \ell}  
    ( \partial_{j} U_i )
    ( \partial_{\ell} U_k )
    \mu_{i j k \ell}(T)
    \nonumber\\ 
    &\quad\quad\quad\quad +
    \frac{ 2 }{ 3 n(\boldsymbol{r}, t) k_B }
    \sum_{i, j}
    \partial_i \left[ \kappa_{i j} \partial_j T(\boldsymbol{r}, t) \right], 
\end{align}
we first note the relation
\begin{align}
    \frac{ d \langle T \rangle }{ d t }
    &=
    \int 
    \frac{ d^3 r }{ N }
    \left[
    n(\boldsymbol{r}, t)
    \frac{ \partial T(\boldsymbol{r}, t) }{ \partial t }
    +
    T(\boldsymbol{r}, t)
    \frac{ \partial n(\boldsymbol{r}, t) }{ \partial t }
    \right] \nonumber\\
    &=
    \left\langle 
    \frac{ \partial T }{ \partial t }
    \right\rangle
    +
    \sum_i
    \frac{ \dot{\sigma}_i(t) }{ \sigma_i(t) }
    \left(
    \frac{ \langle r_i^2 T \rangle }{ \sigma_i^2(t) }
    -
    \langle T \rangle
    \right),
\end{align}
where we utilized the continuity equation. Then multiplying the temperature balance equation by $n(\boldsymbol{r}, t) / N$ and integrating over $d^3 r$ gives 
\begin{align} \label{eq:averaged_T_diffEQ} 
    \frac{ d \langle T \rangle }{ d t }
    &+ 
    \frac{ 5 }{ 3 } 
    \sum_i 
    \frac{ \dot{\sigma}_i(t) }{ \sigma_i(t) }
    \langle T \rangle
    -
    \sum_i
    \frac{ \dot{\sigma}_i(t) }{ \sigma_i(t) }
    \left(
    \frac{ \langle r_i^2 T \rangle }{ \sigma_i^2(t) }
    -
    \langle r_i \partial_i T \rangle
    \right) \nonumber\\
    &\quad = 
    \frac{ 2 }{ 3 N k_B }
    \sum_{i, j, k, \ell} 
    \frac{ \dot{\sigma}_i(t) }{ \sigma_i(t) }
    \delta_{ i, j }
    \left( \int d^3 r \mu_{i j k \ell} \right)
    \delta_{ k, \ell }
    \frac{ \dot{\sigma}_{\ell}(t) }{ \sigma_{\ell}(t) } 
    \nonumber\\ 
    &\quad\quad\quad\quad +
    \frac{ 2 }{ 3 N k_B }
    \sum_{i, j}
    \int d^3 r \left[ \partial_i ( \kappa_{i j} \partial_j T ) \right] . 
\end{align}

Combining equations (\ref{eq:averaged_sigma_diffEQ}) and (\ref{eq:averaged_T_diffEQ}), we get
\begin{align} \label{eq:averaged_Tsigma_diffEQ}
    \frac{ d \langle T \rangle }{ d t }
    &+ 
    \frac{ 5 }{ 3 } 
    \sum_i 
    \frac{ \dot{\sigma}_i(t) }{ \sigma_i(t) }
    \langle T \rangle \nonumber\\
    &\quad -
    \frac{ m }{ k_B }
    \sum_i
    \dot{\sigma}_i(t)
    \left[
    \ddot{\sigma}_i(t)
    +
    \omega_i^2 \sigma_i(t) 
    \right] \nonumber\\
    &\quad \approx 
    \frac{ 2 }{ 3 N k_B }
    \sum_{i, j, k, \ell} 
    \frac{ \dot{\sigma}_i(t) }{ \sigma_i(t) }
    \delta_{ i, j }
    \left( \int d^3 r \mu_{i j k \ell} \right)
    \delta_{ k, \ell }
    \frac{ \dot{\sigma}_{\ell}(t) }{ \sigma_{\ell}(t) } 
    \nonumber\\ 
    &\quad\quad -
    \frac{ 1 }{ N k_B }
    \sum_{i, j, k, \ell} 
    \frac{ \dot{\sigma}_i(t) }{ \sigma_i(t) }
    \left( \int d^3 r r_i \partial_j \mu_{i j k \ell} \right)
    \delta_{k, \ell} 
    \frac{ \dot{\sigma}_k }{ \sigma_{\ell} } \nonumber\\
    &\quad\quad +
    \frac{ 2 }{ 3 N k_B }
    \sum_{i, j}
    \int d^3 r \left[ \partial_i ( \kappa_{i j} \partial_j T ) \right].
\end{align}
At this point, conservation of energy has that 
\begin{align}
    E_{\rm total}
    &=
    \frac{ m }{ 2 }
    \sum_i 
    \left(
    \omega_i^2 \langle r_i^2 \rangle
    +
    \int \frac{ d^3 r d^3 v }{ N } 
    f(\boldsymbol{r}, \boldsymbol{v}, t) v_i^2
    \right) \nonumber\\
    &=
    \frac{ m }{ 2 }
    \sum_i 
    \left(
    \omega_i^2 \sigma_i^2
    + 
    \int \frac{ d^3 r d^3 v }{ N } 
    f(\boldsymbol{r}, \boldsymbol{v}, t) v_i^2
    \right),
\end{align}
where $E_{\rm total}$ is the total energy of the hydrodynamic system. Therefore, the relation above along with Eqs.~(\ref{eq:T_vs_U}) and (\ref{eq:flow_velocity_ansatz}) motivates the form for $\langle T \rangle$ as
\begin{align} \label{eq:average_temperature_ansatz}
    \langle T \rangle  
    &=
    \frac{ 2 E_{\rm total} }{ 3 k_B }
    -
    \frac{ m }{ 3 k_B }
    \sum_i 
    \left[
    \omega_i^2 \sigma_i^2(t) 
    +
    \dot{\sigma}_i^2(t)
    \right],
\end{align}
and its time-derivative 
\begin{align}
    \frac{ d \langle T \rangle }{ d t }
    &=
    -\frac{ 2 m }{ 3 k_B }
    \sum_i 
    \left[
    \omega_i^2 \dot{\sigma}_i(t) \sigma_i(t) 
    +
    \ddot{\sigma}_i(t) \dot{\sigma}_i(t)
    \right]. 
\end{align} 
Plugging these relations into Eq.~(\ref{eq:averaged_Tsigma_diffEQ}) and assuming each axis can be solved independently, we obtain 
\begin{align} \label{eq:sigmas_EOM_unsimplified}
    & \dot{\sigma}_i(t) 
    \left[
    \ddot{\sigma}_i(t)
    +
    \omega_i^2 \sigma_i(t) 
    \right] \nonumber\\
    & + 
    \frac{ \dot{\sigma}_i(t) }{ \sigma_i(t) }
    \left[
    \frac{ 1 }{ 3 }
    \sum_j 
    \left(
    \omega_j^2 \sigma_j^2(t) 
    +
    \dot{\sigma}_j^2(t)
    \right)
    -
    \frac{ 2 E_{\rm total} }{ 3 m }
    \right] \nonumber\\
    &\approx 
    \frac{ 3 }{ 5 N m }
    \sum_{j, k, \ell} 
    \frac{ \dot{\sigma}_i(t) }{ \sigma_i(t) }
    \left( \int d^3 r r_i \partial_j \mu_{i j k \ell} \right)
    \delta_{k, \ell} 
    \frac{ \dot{\sigma}_k }{ \sigma_{\ell} } \nonumber\\
    &\quad -
    \frac{ 2 }{ 5 N m }
    \sum_{j, k, \ell} 
    \frac{ \dot{\sigma}_i(t) }{ \sigma_i(t) }
    \delta_{ i, j }
    \left( \int d^3 r \mu_{i j k \ell} \right)
    \delta_{ k, \ell }
    \frac{ \dot{\sigma}_{\ell}(t) }{ \sigma_{\ell}(t) } \nonumber\\
    &\quad -
    \frac{ 2 }{ 5 N m }
    \sum_{j}
    \int d^3 r \left[ \partial_i ( \kappa_{i j} \partial_j T ) \right].
\end{align}

Finally, the conserved total energy $E_{\rm total}$, is made up of the potential energy and thermal equilibrium temperature $T_0$:
\begin{align} \label{eq:total_energy_relation}
    E_{\rm total}
    &=
    \frac{ 3 }{ 2 } k_B T_0
    +
    \frac{m}{2}
    \sum_i
    \omega_i^2 \sigma_{0,i}^2
    =
    3 k_B T_0,
\end{align}
where we utilized that $\sigma_{0,i} = \sqrt{ { k_B T_0 / m \omega_i^2 } }$.

\section{ Considerations for threshold scattering \label{app:threshold_consideration} }

The analytic results obtained for the viscosities in Sec.~\ref{sec:variational_ansatz_method} are applicable for close to threshold dipolar scattering, which is energy independent \cite{Bohn14_PRA}. However this assumption is only appropriate when the collision energy is much smaller than the characteristic dipole energy $E_{\rm dd} = 16 \pi^2 \epsilon_0^2 \hbar^6 / m^3 d^4$, where $d$ is the electric dipole moment \cite{Bohn09_NJP}. At the same time, the transport coefficients are derived with classical kinetic theory that assumes a nondegenerate sample. Implicit in this formulation is, therefore, that the gas temperature remains well above the Fermi temperature $T_F = \hbar \overline{\omega} (6 N)^{1/3} / k_B$ \cite{Butts97_PRA}. The applicability of our current theory requires that temperature lies in the range $T_F < T \ll E_{\rm dd} / k_B$. 

Furthermore, the derivation above relies on the gas being hydrodynamic, as is characterized by the Knudsen number ${\rm Kn}$.  
The requirements to remain in the regime of validity as formulated in Sec.~\ref{sec:variational_ansatz_method} are summarized as  
\begin{subequations} \label{eq:threshold_requirements}
\begin{align}
    \frac{ \hbar^2 }{ 4 m a_d^2 }
    & \gg 
    k_B T
    > 
    \hbar \overline{\omega} ( 6 N )^{1/3}, \\ 
    N 
    &\gg
    \frac{ 15 \sqrt{ \pi } }{ 4 } \frac{ k_B T }{ m \overline{\omega}^2 a_d^2 }, 
\end{align}
\end{subequations}
which is only ever possible if $a_d / a_{\rm HO} \ll 0.04$,
where $a_d = m d^2 / (8 \pi \epsilon_0 \hbar^2)$ is the dipole length and $a_{\rm HO} = \sqrt{ \hbar / ( m \overline{\omega} ) }$ is the harmonic oscillator length. In heteronuclear alkali dimers, these microwave shielded molecules with $d \sim 1$ D and $m \sim 50$ amu have dipole lengths on the order of $a_{d} \sim 5000 a_0$ to $10,000 a_0$, in units of Bohr radius $a_0$. The necessary trap frequencies to permit threshold scattering above $T_F$ would thus need to be of order $\omega \ll 10$ Hz, which is very weak compared to typical ultracold experiments.  
 
For the parameters in Tab.~\ref{tab:system_parameter}, we find that $k_B T / E_{\rm dd} \approx 28$, implying a more accurate cross section would be that obtained from the semi-classical Eikonal approximation \cite{Glauber55_PR, Glauber59_I, Bohn09_NJP}. We opt to proceed with the effective cross section obtained with threshold energy scattering as it still serves to illustrates the effectiveness of our theory, as formulated for arbitrary cross sections.

\section{ A simple functional form for the hydrodynamic volume \label{app:HD_volume_ansatz} }

From Fig.~\ref{fig:ETAvsN_cigar}, we saw that $\eta$ is mostly independent of $\lambda$, which leaves us with $\eta = \eta( N, \Theta )$. Then assuming that $\eta$ is separable in its 2 arguments, this allows us to write $\eta( N, \Theta ) = \eta_N( N ) \eta_{\Theta}( \Theta )$. 
Within the range of $N$ we explore, we could Taylor expand $\eta_N$ around a number of molecules that is sure to be hydrodynamic $N_0$, so that
\begin{align}
    \eta( N, \Theta )
    &\approx 
    \left(
    \eta_N (N_0)
    +
    ( N - N_0 )
    \left. \frac{ \partial \eta_N }{ \partial N } \right|_{N_0}
    \right)
    \eta_{\Theta}( \Theta ). 
\end{align}
Then also assuming that the dependence of $\eta_{\Theta}$ on $\Theta$ purely arises through $\sigma_{\rm coll}(\Theta)$ (i.e. $\eta_{\Theta} = \eta_{\Theta}( \sigma_{\rm coll} )$), we then treat $\xi = \sigma_{\rm coll} / \overline{\sigma}_{\rm coll}$ as a small parameters and Taylor expand $\eta_{\Theta}$ to give
\begin{align}
    \eta( N, \Theta )
    &\approx 
    a 
    +
    b 
    \left( \frac{ N }{ 10^5 } \right)
    \left[
    1
    +
    c \left( \frac{ \sigma_{\rm coll}(\Theta) }{ \overline{\sigma}_{\rm coll} } \right)
    \right],
\end{align}
as in Eq.~(\ref{eq:eta_ansatz}), where 
\begin{subequations}
\begin{align}
    a 
    &=
    \eta_{\Theta}(0) 
    \left(
    \eta_N (N_0)
    -
    N_0
    \left. \frac{ \partial \eta_N }{ \partial N } \right|_{N_0}
    \right), \\
    b
    &=
    10^5 \times 
    \eta_{\Theta}(0)
    \left. \frac{ \partial \eta_N }{ \partial N } \right|_{N_0}, \\
    c
    &=
    \frac{ 1 }{ \eta_{\Theta}(0) }
    \left. \frac{ \partial \eta_{\Theta} }{ \partial \xi } \right|_{\xi = 0},
\end{align}
\end{subequations}
having used the notation $\eta_{\Theta}(0) = \eta_{\Theta}( \xi = 0 )$.

\nocite{*}
\bibliography{main.bib} 

\end{document}